\documentclass[%
aip,
amsmath,amssymb,
reprint,%
]{revtex4-2}
\usepackage{graphicx}
\usepackage{dcolumn}
\usepackage{mathrsfs}
\usepackage{bm}
\usepackage{lineno}
\usepackage{hyperref}
\usepackage{bbm}
\usepackage{ulem}
\usepackage{xcolor}

\begin{document}
\title{Phase Dependent Hanbury-Brown and Twiss effect}

\author{Xuan Tang*}
\affiliation{Department of Physics, City University of Hong Kong, 83 Tat Chee Avenue, Kowloon, Hong Kong, P. R. China}

\author{Yunxiao Zhang*}
\affiliation{College of Precision Instrument and Opto-Electronics Engineering, The State Key Laboratory of Precision Measurement Technology and Instruments, Tianjin University, Tianjin 300072, P. R. China}

\author{Xueshi Guo}
\affiliation{College of Precision Instrument and Opto-Electronics Engineering, The State Key Laboratory of Precision Measurement Technology and Instruments, Tianjin University, Tianjin 300072, P. R. China}

\author{Liang Cui}
\affiliation{College of Precision Instrument and Opto-Electronics Engineering, The State Key Laboratory of Precision Measurement Technology and Instruments, Tianjin University, Tianjin 300072, P. R. China}

\author{Xiaoying Li}
 \email{xiaoyingli@tju.edu.cn}
\affiliation{College of Precision Instrument and Opto-Electronics Engineering, The State Key Laboratory of Precision Measurement Technology and Instruments, Tianjin University, Tianjin 300072, P. R. China}

\author{Z. Y. Ou}
 \email{jeffou@cityu.edu.hk}
\affiliation{Department of Physics, City University of Hong Kong, 83 Tat Chee Avenue, Kowloon, Hong Kong, P. R. China}

\begin{abstract}
* equal contributions in alphabetic order\\

\noindent Hanbury-Brown and Twiss (HBT) effect is the foundation for stellar intensity interferometry. However, it is a phase insensitive two-photon interference effect. Here we extend the HBT interferometer by mixing a reference field with the input fields before intensity correlation measurement. {With the freely available coherent state serving as the reference field, we experimentally demonstrate the phase sensitive two-photon interference effect when the input fields are  thermal fields in { either continuous wave or non-stationary pulsed wave and measure the complete complex second-order coherence function of the input fields without bringing them together from separate locations.} Moreover, we discuss how to improve the signal level by using the more realistic continuous wave broadband anti-bunched light fields as the reference field. Our investigations pave the way for developing new technology of remote sensing and coherent imaging. }

\end{abstract}

\maketitle

Hanbury-Brown and Twiss (HBT) effect \cite{hbt} was the first to reveal intensity fluctuation of an optical field and laid the foundation for the modern quantum optics. Soon after its discovery, it was applied to stellar intensity interferometry of high resolution to measure the size of main sequence stars \cite{hbt-sii} and has received more attention recently \cite{V-sii}. The underlying physical principle of the effect is two-photon interference \cite{fano,glau}, which shows correlations in the fourth-order field quantities.

However, HBT effect is phase-insensitive two-photon effect even though it is an interference phenomenon. Although it is related to the absolute value of the second-order coherence function for thermal light fields, it cannot measure the phase of the second-order coherence function and thus has limited application as compared to the more traditional stellar interferometry based on direct interference of celestial light by Michelson \cite{mich}, which, on the other hand,  has its own problem of limited range because it needs to bring light together for interference \cite{moni}.

 {Recently, following the proposal of long baseline telescope using {nonlocally entangled single-photon state \cite{gottes}, experiment \cite{sm23} was performed with heralded single-photon entangled state in pulsed mode as references} to realize optical interferometric imaging of weak thermal light sources \cite{sm23}. Moreover, a variation of the long baseline interferometry { by merging two separated signal and reference fields into one for holography was proposed and demonstrated}, where the signal wavefront in both intensity and phase was reconstructed from the intensity correlation measurement \cite{sussman}. In these proof-of-principle experiments, quantum states in pulsed mode were used so that it is not applicable to the objects emitting continuous wave light and the involvement of delicate quantum state { makes it less practical because of its sensitivity to losses in long distant distribution}. On the other hand, it is well-known that coherent state, at low average photon number, can be regarded as an entangled single-photon state but indeterministically due to vacuum contribution \cite{gottes}. Therefore,  quantum states are not necessary for realizing the phase dependent intensity correlation.}

{In this paper, we study the HBT interferometer modified by introducing a reference field to realize phase dependent intensity correlation. We perform experiment by using coherent state as reference field due to its easy accessibility. To demonstrate the broad range of applicability of our scheme, we conduct experiments by using thermal fields  in the form of both continuous wave (CW) and pulsed mode as input. Moreover, we show the distance between two separated measurement locations can be far beyond the coherence length \cite{njp,kim,ou22}, which implies potential  resolution increase by enlarging the baseline. 

Notice that our scheme is similar to the previous heterodyne detection technique \cite{joh} for realizing phase-sensitive measurement of celestial light, which mixes the thermal light fields with {\it strong} coherent states as local oscillators. However, heterodyne detection technique suffers the problem of shot noise at low input photon level \cite{isi}. Our scheme is not limited by shot noise problem because we use {\it weak} coherent states} and measure intensity correlation for which vacuum has no contribution. Furthermore, for the case of very weak input field, we discuss the cost of using coherent states as reference light by comparing our scheme with the recently proposed entanglement-based telescopy scheme \cite{gottes} and propose to improve the signal-to-noise ratio by exploiting more realistic continuous wave broadband anti-bunched light fields as the reference fields. Our investigation not only is useful for astronomical applications, but is also valuable for developing new technique of synthetic aperture imaging of distant objects, which are either actively or passively illuminated  \cite{sensing,sensing2}.

\begin{figure}[t]
\includegraphics[width=8.5cm]{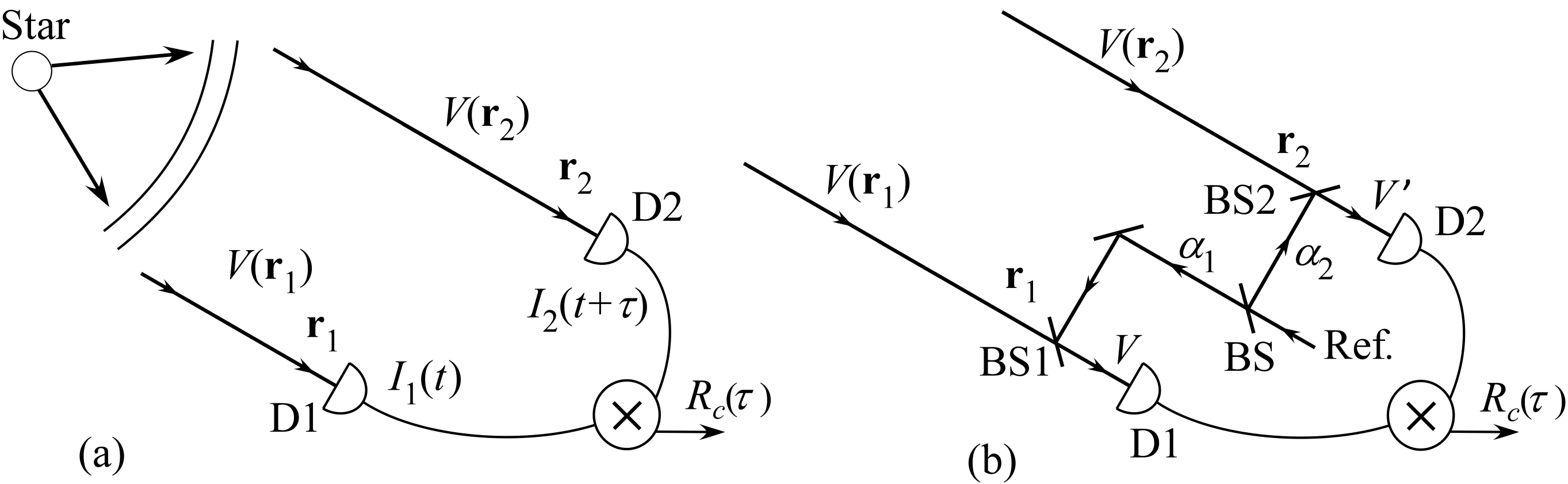}
	\caption{(a) Schematic of the Hanbury Brown and Twiss experiment for intensity correlations observed at two points. (b) Two-photon interference scheme with a reference light introduced for retrieving second-order coherence function $\gamma({\bf r}_1,{\bf r}_2, \tau)$ between input fields $V({\bf r}_1), V({\bf r}_2)$.  D1, D2: photo-detectors, BS, beam splitter.}
	\label{sch}
\end{figure}

{Figure 1(a) sketches the scheme of Hanbury-Brown and Twiss (HBT) experiment for traditional intensity correlations measurement\cite{hbt, hbt-sii}. Denoting the incoming fields at two locations as  $V({\bf r}_1)$, $V({\bf r}_2)$ and assuming they are of thermal nature, the intensity correlation is associated with the absolute value of second-order coherence function of the stellar optical field  $\gamma(\tau) \equiv \gamma({\bf r}_1,{\bf r}_2, \tau) =  \langle V({\bf r}_1, t)V^*({\bf r}_2, t+\tau)\rangle$ through the relation  $\langle I_1(t) I_2(t+\tau) \rangle = \langle I_1(t)\rangle \langle I_2(t+\tau)\rangle (1+|\gamma(\tau)|^2)$, where $I_i=|V({\bf r}_i)|^2$ ($i=1,2$) is the intensity measured at $ {\bf r}_i$. Fig. 1(b) shows the modified HBT interferometer, which is realized by introducing reference fields and mixing them with input fields $V({\bf r}_1)$, $V({\bf r}_2)$ to achieve phase sensitive two-photon interference. The reference fields are in coherent states when they come from the splitting of a laser source by a beam splitter (BS). The input fields $V({\bf r}_1)$, $V({\bf r}_2)$ are mixed with two coherent fields ($\alpha_{1},\alpha_{2}$) by BS1 and BS2. Detectors D1 and D2 placed at the output ports of BS1 and BS2 respectively measure the intensities of mixed fields $V$ and $V'$. Assuming both the input and reference fields are single mode continuous wave (CW), }
it is straightforward to find the result of intensity correlation or coincidence as \cite{ou22}
\begin{eqnarray}\label{As2}
R_c(\tau) &\propto & \Gamma_{2,2}(\tau)  \equiv  \langle |V(t)|^2|V'(t+\tau)|^2\rangle\cr & \propto &
I_1I_2 [1+\lambda(\tau)] + |\alpha_1\alpha_2|^2   +(I_1|\alpha_2|^2 + I_2|\alpha_1|^2)\cr &&\hskip 0.5 in \times [1+ \xi|\gamma(\tau)|\cos(\varphi_{\gamma}+\Delta\phi_{\alpha})],~~~~
\end{eqnarray}
{where $I_i\equiv \langle |V({\bf r}_i,t)|^2 \rangle (i=1,2)$, $|\gamma(\tau)|$ and $\varphi_{\gamma}$ are the magnitude and phase of $\gamma(\tau)$, $\Delta\phi_{\alpha} \equiv \phi_{\alpha_2}-\phi_{\alpha_1}$ is phase difference between reference beams $\alpha_{1},\alpha_{2}$, and $1+ \lambda(\tau) \equiv  \langle |V({\bf r}_1,t)|^2 |V({\bf r}_2, t+\tau)|^2\rangle/ I_1 I_2$ is the normalized intensity correlation function of the input fields. The coefficient $\xi$ is determined by the intensities of input and reference fields and by the mode matching. It equals $\xi \equiv 2|\alpha_1\alpha_2|\sqrt{I_1I_2}/(I_1|\alpha_2|^2 + I_2|\alpha_1|^2)$ when the input and reference fields are ideally mode-matched}. Note that the phase $\varphi_{\gamma}$ varies with distance and orientation of two locations $ {\bf r}_1,{\bf r}_2$. In deriving Eq.(\ref{As2}), we assume $\alpha_{1,2}$ has stable phases and the phase between input field $V({\bf r},t)$ and reference light randomly fluctuates so that each detector exhibits no interference. Normally, two input fields from a star have identical intensity $I_1=I_2$ and are of thermal nature, so $\lambda(\tau) = |\gamma(\tau)|^2$. Setting $|\alpha_1|^2=|\alpha_2|^2=I_1=I_2\equiv I$ in Eq.(\ref{As2}), we have
\begin{eqnarray}\label{As3}
R_c(\tau) \propto I^2(4+|\gamma|^2)[1+ {\cal V}\cos(\varphi_{\gamma}+\Delta\phi_{\alpha})],~~~~~~~
\end{eqnarray}
where ${\cal V} \equiv 2\xi|\gamma(\tau)|/(4+|\gamma(\tau)|^2)$ is the visibility of interference fringe  with $\xi$ accounting for non-ideal mode match. It has a maximum value of 40\% when $|\gamma(\tau)|=1$ and $\xi=1$. Notice that the phase information of $\gamma(\tau)$ is in the fringe pattern and the absolute value $|\gamma(\tau)|$ can be extracted from visibility ${\cal V}$. Different from Fig. 1(a), which can measure $|\gamma(\tau)|$ and find the angular diameter of stars, here we can measure the complete complex function of $\gamma(\tau)$, from which the intensity distribution of the star can be extracted. Moreover, knowledge of $\gamma({\bf r}_1,{\bf r}_2, \tau)$ for a large separation of $ {\bf r}_1,{\bf r}_2$ will lead to high angular resolution by a Fourier transformation \cite{sensing}.

\begin{figure}[t]
\includegraphics[width=8.5cm]{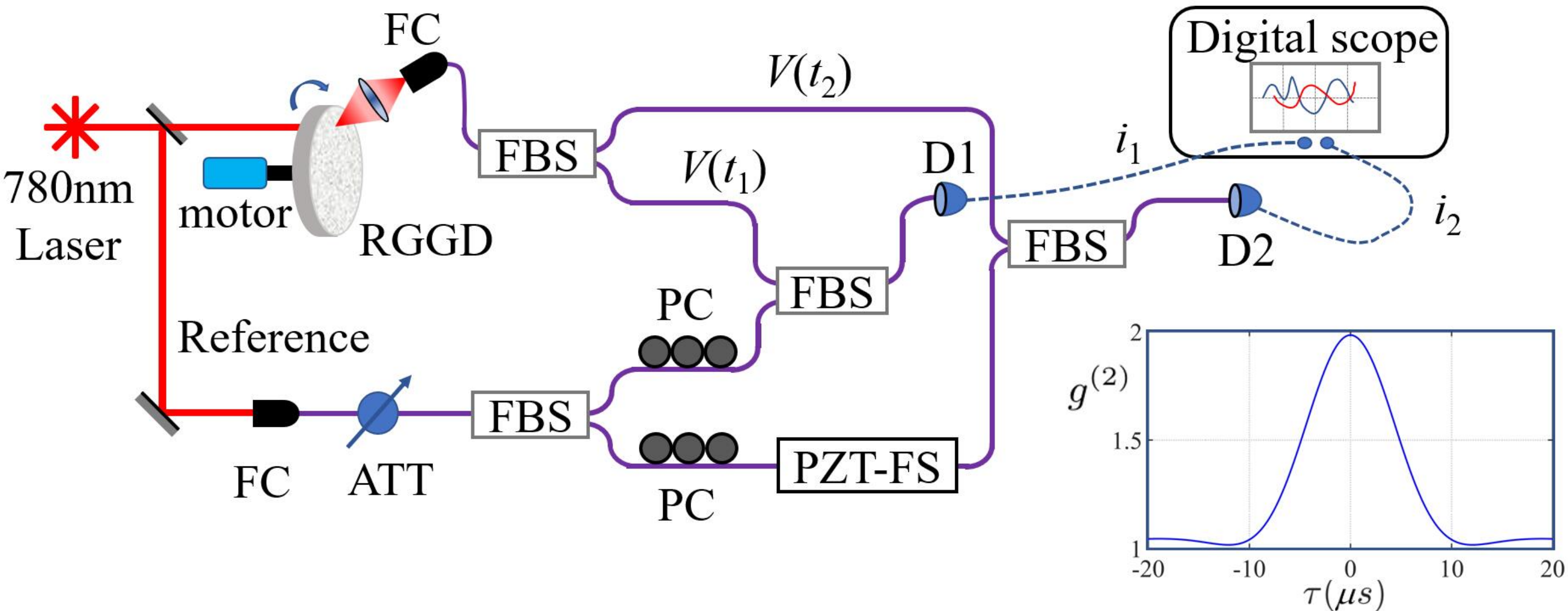}
	\caption{Experimental arrangement for a quasi-cw thermal light field. RGGD: rotating ground glass disc; PC: polarization control; PZT-FS:  piezoelectric transducer-driven fiber stretcher; FC: fiber coupler; FBS: fiber beam splitter; D1, D2: photo-detectors; ATT: attenuator.  Inset: $g^{(2)}(\tau)$ for the input field when the reference field from laser is blocked.}
	\label{ex2}
\end{figure}

We implement the scheme in Fig.\ref{sch} (b) in a proof-of-principle experiment with both continuous wave (CW) and pulsed thermal fields.  The experiment with CW thermal light fields perhaps better mimics the situation in astronomy where celestial light is continuous and of thermal nature. The schematics is shown in Fig.\ref{ex2} where thermal light fields are the scattered light of a coherent field from a single-frequency laser  by a rotating ground glass disk (RGGD). We couple the light into a single-mode fiber with a fiber coupler (FC) for best spatial mode match. The coupled light is then split into two by a fiber beam splitter (FBS) and sent to two separate locations ($V_1(t), V_2(t)$), emulating the celestial light collected at two locations. The thermal light fields at the two locations are respectively mixed with coherent fields ($\alpha_1,\alpha_2$) which are also from the splitting of the laser as reference. The intensity of the coherent fields is reduced by attenuation (ATT) to match that of the thermal fields (about 10 $\mu W$ each) and are adjusted in polarization by polarization controllers (PC). The phase difference of the coherent states is scanned by a piezoelectric transducer-driven fiber stretcher (PZT-FS). Two fast photo-detectors (D1, D2, Thorlab PDB450A, 150 MHz bandwidth) are used to record the photo-currents $i_1(t), i_2(t)$ of the mixed fields at two locations (When detectors' response time is faster than field fluctuation time, the  input fields can be viewed as a single-mode field). The photo-currents are then collected by a digital oscilloscope for data processing.

We first measure $g^{(2)}(\tau) \equiv \langle i_1(t) i_2(t+\tau)\rangle/\langle i_1(t)\rangle \langle i_2(t)\rangle$ on the thermal fields directly  by blocking the reference light, where $\langle ...\rangle$ is the time average over $\Delta T=2 ms$. The measured $g^{(2)}(\tau)$ is plotted as the inset of Fig.\ref{ex2}. According to the relation  $g^{(2)}(\tau)=1+|\gamma(\tau)|^2$, we find that the coherence time of the thermal field is about 7 $\mu s$, which is determined by the grain size and the rotating speed of the ground glass disc.


\begin{figure}[t]
\includegraphics[width=8.5cm]{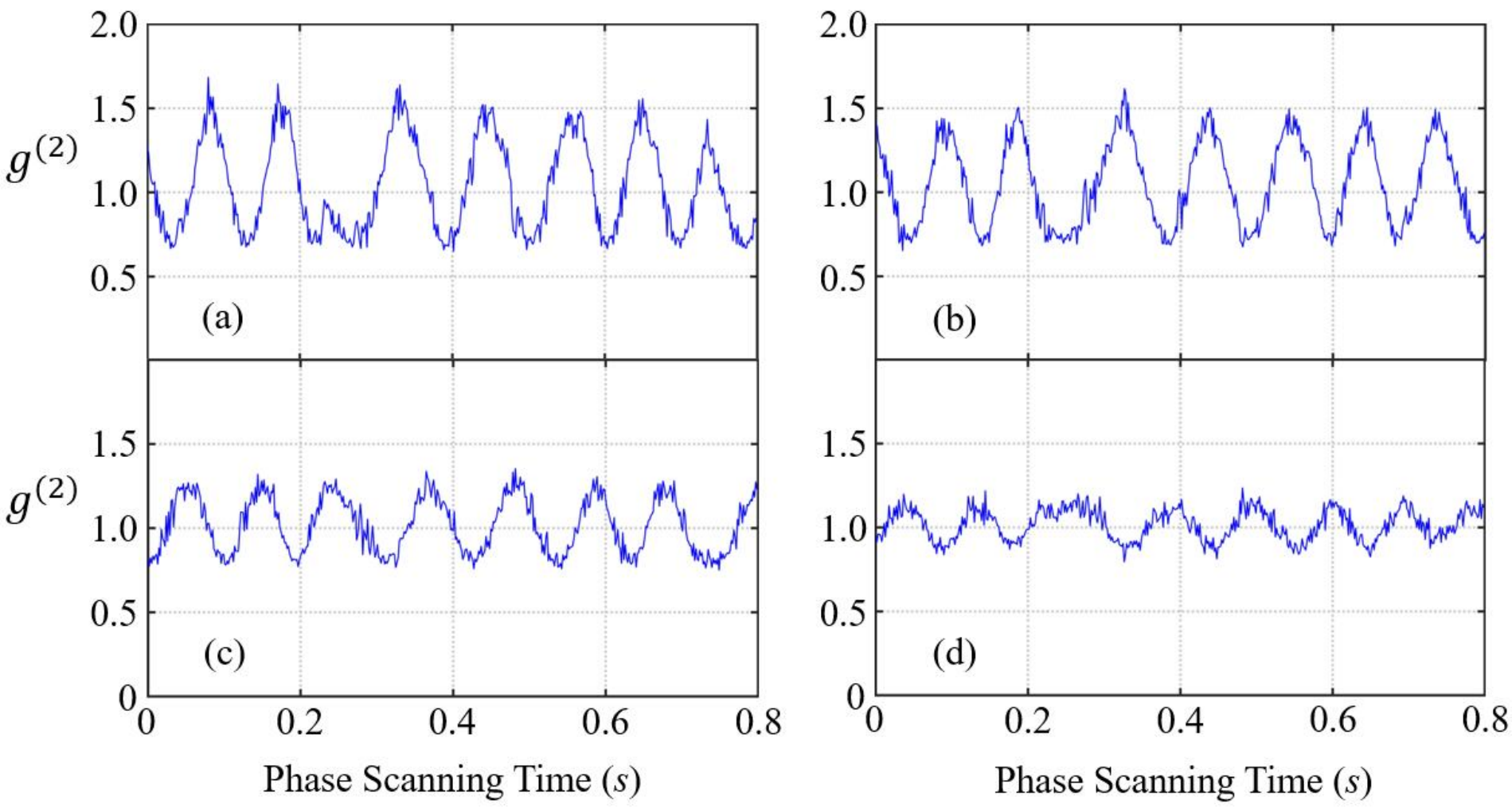}
	\caption{Interference fringes shown in normalized intensity correlation function $g^{(2)}(\tau)$ as a function of phase at different time delays: (a) $\tau = - 0.5 \mu s$; (b) $\tau = -3.0 \mu s$; (c) $\tau = 6.5 \mu s$; (d) $\tau = 9.5 \mu s$.}
	\label{fringe-cw}
\end{figure}

\begin{figure}[t]
\includegraphics[width=8cm]{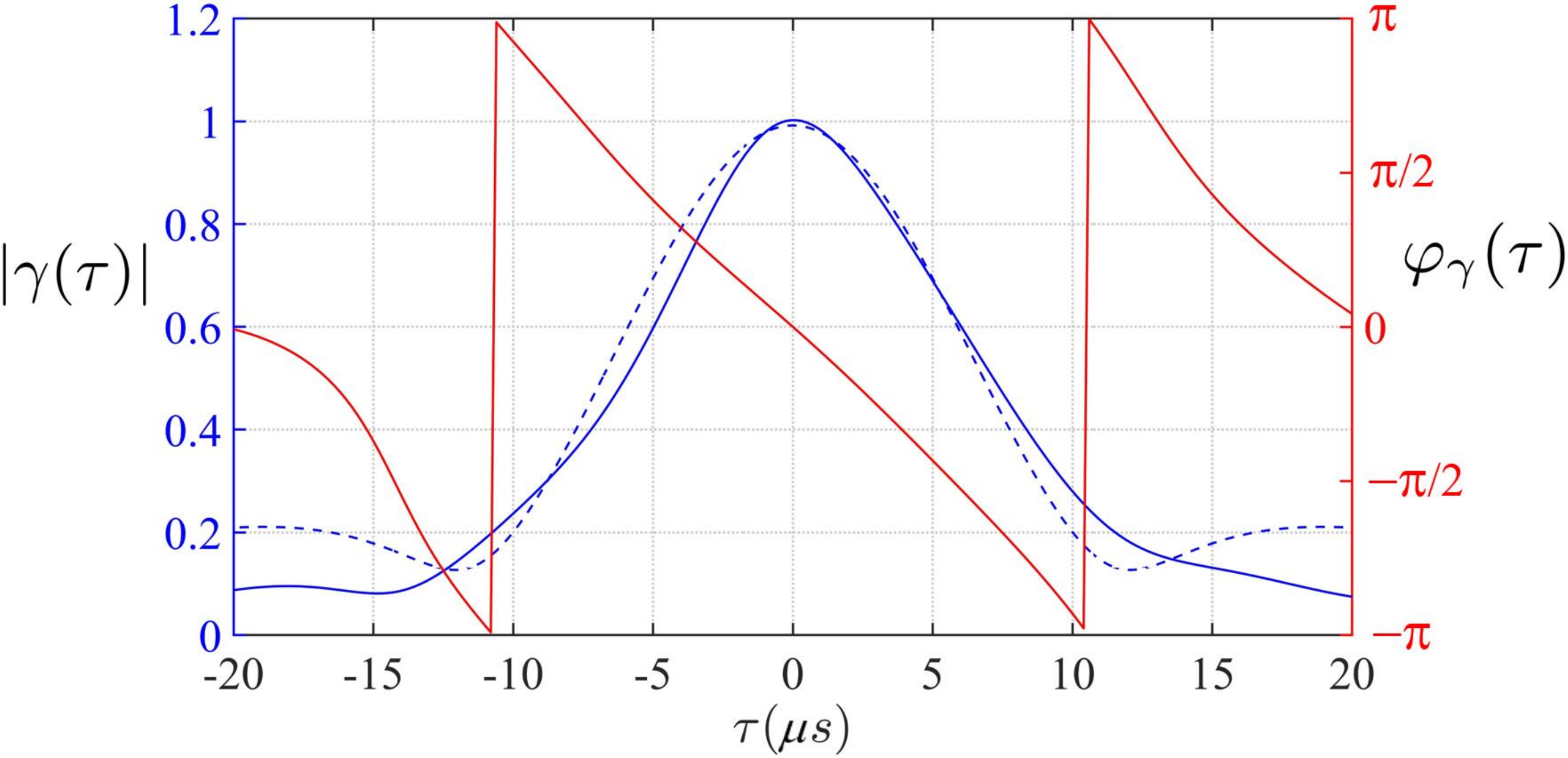}
	\caption{Extracted $|\gamma(\tau)|$ (blue) and $\varphi_{\gamma}(\tau)$ (red) as a function of delay $\tau$. Dashed line (black) is the $|\gamma(\tau)|$-value derived from $g^{(2)} = 1+|\gamma(\tau)|^2$ by directly measuring $g^{(2)}$ without the coherent fields. }
	\label{V-cw}
\end{figure}

When the reference fields are unblocked and their relative phase is scanned by varying the voltage of piezoelectric transducer (PZT) mounted on fiber stretcher (FS),
$g^{(2)}(\tau)$ is again measured from the recorded photo-currents $i_1(t), i_2(t)$. The results for different delay $\tau$ show interference fringe in Fig.\ref{fringe-cw}. Visibility is measured from each fringe. The maximum visibility of 35\% is observed near $\tau =0$. The deviation of the maximum visibility from the theoretical value of 40\% may come from polarization mode mismatch of the fields. After considering this, we can extract $|\gamma|$ from visibility by using ${\cal V} = 2\xi |\gamma|/(4+|\gamma|^2)$ with $\xi=0.35/0.4=0.875$. 

It is worth noting that the fringes in Fig.\ref{fringe-cw} are shifted for different $\tau$. We can extract the phase shift relative to that of $\tau=0$ ($\varphi_{\gamma}(0)=0$ because $\gamma(0)=1$). Both $|\gamma|$ and $\varphi_{\gamma}$ are plotted as a function of $\tau$ in Fig.\ref{V-cw}. Notice that $\varphi_{\gamma}$ depends nearly linearly on $\tau$: $\varphi_{\gamma}\approx -\Delta\Omega_{D} \tau$ ($\Delta\Omega_D=3.1\times 10^5rad/s$), which is due to a Doppler shift $\Delta\Omega_{D}$ from the center frequency of the laser for the scattered quasi-thermal light by the rotation (moving) of the ground glass. Therefore, we are able to measure the complete complex function of $\gamma(\tau)$ with this technique. The value of $|\gamma|$ can also be extracted directly from the measured $g^{(2)}(\tau) = 1 + |\gamma(\tau)|^2$ of thermal fields (the inset in Fig. \ref{ex2}), which is represented by the dashed curve in Fig.\ref{V-cw} and is consistent with the blue one indirectly obtained from visibility of interference fringes at different time delay (see Fig. \ref{fringe-cw}).

In the experiment described above, we use multiplication of photo-currents as intensity correlation or coincidence measurement method. It works for input fields with intensity at a relative high level. For detectors with a bandwidth of about 150 MHz and noise equivalent power of 7.5 pW/$\sqrt{Hz}$, the intensity of input fields must be higher than $10^{12}$ photons per second in order to have SNR greater than 1. Hence, for the input fields with low brightness, electronic noise will overwhelm the light signal. In this case, we need to resort to photon counting method by using single-photon detectors to effectively extract light signal. Instead of switching to single-photon detectors and repeating the experiment above, we apply photon counting technique to the pulsed field case.

\begin{figure}[t]
\includegraphics[width=8cm]{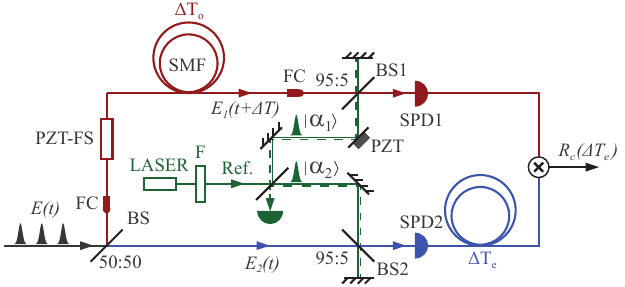}
	\caption{Experimental arrangement for pulsed thermal light. BS: beam splitter; F: filter; SPD: single photon detector; SMF: single mode fiber; PZT: piezoelectric transducer.}
	\label{ex}
\end{figure}

We then implement the interference scheme in Fig.\ref{sch} (b) with a train of pulsed thermal field as input, which can mimic the situation of distant objects being actively illuminated by a pulsed laser  in the applications of remote sensing and LIDA.
The schematics is shown in Fig.\ref{ex}. The input thermal field $E(t)$ with a pulse duration of about 9 ps is originated from the individual signal field of spontaneous four-wave mixing in pulse pumped single mode fiber \cite{Ma-pra11}. After splitting the field with a 50/50 BS, two fields $E_1(t)$ and $E_2(t)$ are then sent to different locations and mixed with weak coherent fields ($\alpha_1,\alpha_2$) by using BS1 and BS2, respectively. The coherent fields are obtained by passing the output of a mode-locked fiber laser at $1.55\mu m$ wavelength through a filter (F) and then splitting it by a 50/50 BS. The repetition rate of the mode-locked laser is about 50 MHz, which is synchronized with that of the input thermal field. The spectrum of filter F is chosen to well-match that of the thermal fields. The optical path lengths of the thermal fields of $E_1(t+\Delta T_o)$ and $E_2(t)$ are different, where $\Delta T_o=21 ns$ is the delay introduced by sending $E_1$ field through an extra piece of single-mode fiber (SMF). This arrangement implies that the separation of two location $ {\bf r}_1,{\bf r}_2$ in Fig. 1(b) is far beyond the coherence length of input fields. In the experiment, the polarization and temporal modes of two fields mixed at BS1 and BS2 are well-matched, and the output of BS1 and BS2 are respectively detected by single photon detectors (SPD1 and SPD2). The two SPDs (InGaAs-based) are operated in a gated Geiger mode. The 2.5-ns gate pulses coincide with the arrival of photons at SPDs. In the process of observing the interference through the coincidence of two SPDs,  the phase difference of two input thermal fields ($E_1(t+\Delta T_o),E_2(t)$) is scanned by using a piezoelectric transducer (PZT) mounted on a mirror, and an electronic delay $\Delta T_e = \Delta T_o$ is introduced to the output of SPD2.


The theory for the pulsed interference scheme in Fig.\ref{ex} was outlined in Ref.\citenum{ou22}. We can model the input pulse train of pulse separation $T_p$ as $E(t) = \sum_j A_j g(t-j T_p)$ and the coherent state of reference light as $E_{ref}(t) = \alpha \sum_j  f(t-j T_p)$ with $g(t), f(t)$ being the normalized mode functions ($\int dt g^2(t)=1 = \int dt f^2(t)$) for the pulses and $A_j, \alpha$ as the amplitude of the j-th pulse of the two fields, respectively. The input fields and the reference fields are assumed to be matched pulse by pulse.  To emulate fields at different locations, we introduce an optical delay $\Delta T_o$ for one of the fields, say $E_1$. The corresponding reference field ($\alpha_1$) also needs proper delay for pulse overlap. Denote $\langle |A_j|^2\rangle_j \equiv \bar n$ and $|\alpha|^2$ as the average photon number per pulse for the input and reference fields, respectively. Assuming BS1 and BS2 used to mix the input and reference fields are 50:50 beam splitters, it is straightforward to find the coincidence rate as \cite{ou22}
\begin{eqnarray}\label{As-p}
R_c(\Delta T_e) &=& \frac{1}{4}R_p \Big\{ \bar n^2g^{(2)} + |\alpha|^4 + 2\bar n|\alpha|^2 \cr
&&\hskip 0.1 in \times [1+ \beta_1\beta_2|\gamma(\Delta N)|\cos(\varphi_{\gamma}+\Delta\phi_{\alpha})]\Big\}.~~~~
\end{eqnarray}
Here, $\Delta T_e$ is the electronic delay used to compensate the optical delay $\Delta T_o$ and $R_p\equiv 1/T_p$ is the repetition rate of the pulse train. $\gamma(\Delta N) =|\gamma|e^{i\varphi_{\gamma}} \equiv \langle A_j A_{j+\Delta N}^*\rangle_j/\langle |A_j|^2 \rangle_j$ with $\Delta N\equiv$  integer part of $(\Delta T_o -\Delta T_e)/T_R (T_R$ is the coincidence window and is assumed to be smaller than $T_p$) describes the coherence property of the input field, similar to the cw case. $g^{(2)} \equiv \langle |A_j|^2 |A_{j+\Delta N}|^2\rangle_j/\langle |A_j|^2 \rangle_j^2$ is the normalized intensity correlation. $\beta_{1,2} \equiv \int dt f(t) g_{1,2}(t)$ are the temporal mode matching coefficients for the two fields split from the input field. We give them different subscript label because they may go through different media and suffer different distortion when used for sensing or ranging.  In our experiment, the coherence time of the thermal field is three orders of magnitude smaller than the pulse duration, so that $|\gamma(\Delta N)|=0$ unless $\Delta N = 0$, which is satisfied when we set $\Delta T_o= \Delta T_e$. Moreover, it is worth noting that the expressions of Eq. (\ref{As-p}) and Eq.(\ref{As3}) are actually the same if we replace $R_p \bar n$, $|\alpha|^2$ and $\beta_1\beta_2$ in Eq.(\ref{As-p}) with $I$($=I_{1,2}=|\alpha_1|^2=|\alpha_2|^2$) and $\xi$, respectively.

\begin{figure}[t]
\includegraphics[width=7cm]{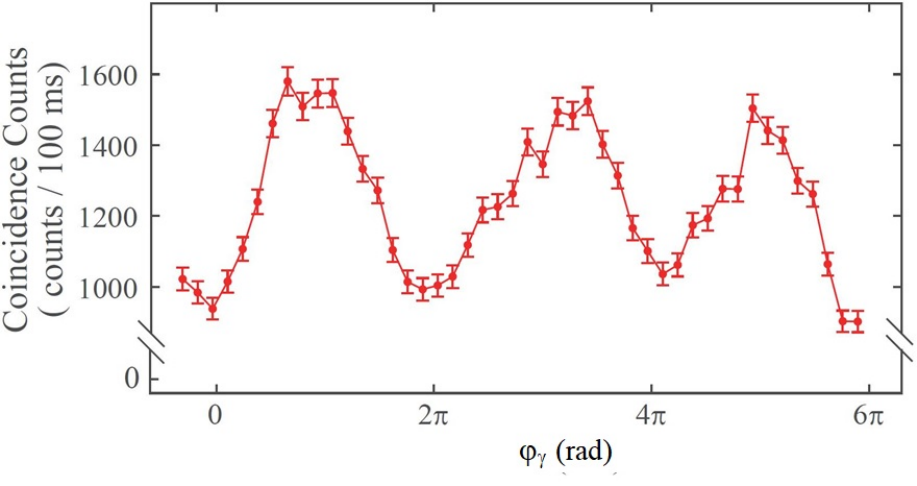}
	\caption{ Interference fringe seen as coincidence counts when the relative phase between the two inputs $E_1(t+\Delta T_o),E_2(t)$ is scanned. In the experiment, delay $\Delta T_o$ is three orders of magnitude larger than the coherence time of input $E(t)$.} 
	\label{data}
\end{figure}

Figure \ref{data} plots the coincidence rate obtained by applying a ramp voltage on the PZT to  scan the phase difference $\Delta \varphi_{\gamma}$ between two thermal input fields. An interference pattern is observed. During the measurement, the detection rate of each SPD is about 0.03 photons/pulse, and the single count rate of each SPD stays constant because there is no phase relation between the thermal and coherent fields. Note that the phase difference $\Delta \varphi_{\alpha}$ between two weak coherent fields ($\alpha_1,\alpha_2$) fluctuates with the environment, causing irregular interference fringes in Fig.\ref{data}. In a real application, we can lock this phase by sending back the coherent fields to form a Michelson interferometer (see the dashed lines in Fig.\ref{ex}). The visibility of interference in Fig.\ref{data} is about 22\%, which deviates from the predicted maximum value of 40\%. We believe this is caused by non-ideal mode match between the thermal state and coherent state. The coherent state obtained by attenuating and filtering the mode-locked fiber laser is in a single temporal mode. However, from the measured normalized intensity correlation function $g^{(2)}=1.65$ of the thermal field, which is obtained by blocking the coherent fields, we can deduce that the mode number of thermal field is about 1.5, leading to mode mis-match with $\beta_{1,2} < 1$.


Because the current method depends on coincidence measurement whose rate is proportional to the square of the incoming photon rate, it is similar to stellar intensity interferometry method \cite{hbt-sii} and suffers low signal level for dim light, in contrast to the scheme with entangled single-photon states as the local oscillators\cite{gottes} and the direct interference scheme in Michelson's stellar interferometry \cite{mich}, where the signal is proportional to the celestial photon rate. The reason is the second term of $|\alpha_1\alpha_2|^2$ in Eq.(\ref{As2}), which requires the matching of the photon numbers between the thermal states and coherent states for the maximum visibility of interference. This term comes from two-photon contributions of coherent states but is zero if we use entangled single-photon states to replace the coherent states as the reference light. Without this term, we can increase $\alpha$ to enhance the interference part, in a similar way as homodyne detection where local oscillators act as an amplifier to increase the optical signal.
One method to eliminate the term of $|\alpha_1\alpha_2|^2$ is to use a modified coherent state where the two-photon term is canceled by destructive two-photon interference with a two-photon source \cite{jap93,vy93,lu02,zhu05}. But this still applies to the case of $|\alpha|^2 \ll 1$ since three-photon term of coherent states can contribute to coincidence rate.

Of course, single-photon states as reference will not have the aforementioned issue  as suggested in the original proposal \cite{gottes,sm23}. In practice, a perfectly anti-bunched light fields such as those from single emitters \cite{kim77,wal87} will do the work. But a field is in a single-photon state only within a certain time window $T_{SPS}$, which ideally is simply $1/I_{SPS}$ with $I_{SPS}$ as the average photon rate of the single-photon field ($T_{SPS}$ is viewed as the average time separation between photons). Let us estimate the signal rate for the scheme with this type of field. In general,  we can write a normalized two-photon correlation function as $g^{(2)}_{Ref} = 1 + \lambda_{Ref}(\tau)$ for the reference field. A perfectly anti-bunched light field as the reference will have $\lambda_{Ref}(\tau) = -1$ for $|\tau|< T_{Ref}\equiv T_{SPS}$ \cite{kim77}. Then, it is straightforward to show that Eq.(\ref{As2}) becomes
\begin{eqnarray}\label{Anti}
&&\Gamma_{2,2}(\tau)  =
I_1 I_2 [1+\lambda(\tau)] + |\alpha_1\alpha_2|^2[1+\lambda_{Ref}(\tau)]  \cr && \hskip 0.2in + (I_1|\alpha_2|^2+I_2|\alpha_1|^2)[1+ \xi|\gamma(\tau)|\cos(\varphi_{\gamma}+\Delta\phi_{\alpha})],\cr &&
\end{eqnarray}
where $|\alpha_{1,2}|^2 \equiv I_{Ref1,2}$ are the intensities of the two anti-bunched reference fields. Coincidence rate, or the signal rate is an integration over the coincidence window of $T_R$: $R_c(\tau_e) = \int_{[T_R]}\Gamma_{2,2} (\tau+\tau_e) d\tau \approx \Gamma_{2,2}(\tau_e)T_R$ for time resolved coincidence measurement at a time delay of $\tau_e$ if $T_R\ll T_c$ ($T_c$ is the coherence time of the incoming thermal fields). If we can manage $\lambda_{Ref}(\tau_e) = -1$ for $\tau_e < T_{Ref}$ but with $T_{Ref} {\gtrsim} T_c$ (for complete $\gamma(\tau)$ measurement), then the second term is zero and the first term is negligible if we make single-photon field much larger than the incoming thermal field or $I_{Ref}\equiv |\alpha_{1,2}|^2 \gg I_{1,2} \equiv I$. We now can measure $\gamma(\tau)$ with a signal rate of $I I_{Ref}T_R = \zeta I$, which is proportional to the photon rate $I$ of the incoming light field with $\zeta\equiv I_{Ref} T_R$. But as we said earlier, $I_{Ref}$ is at most $1/T_{Ref}$ so $\zeta\approx T_R/T_{Ref} \ll T_c/T_{Ref} {\lesssim} 1$. We can eliminate the requirement of $T_{Ref} > T_c$ if we can add adjustable optical delay $T_o \sim \tau_e$ between the two reference fields so that $\lambda_{Ref}(\tau_e-T_o)= -1$ when $\tau_e$ is within the coincidence window of $T_R$. This allows $\tau_e$ cover the range of $\gamma(\tau)$ for a complete measurement of $\gamma(\tau)$ function. Then this only needs $T_{Ref}\sim T_R$, which is equivalent to temporal mode match in pulse case, so that $\zeta\approx T_R/T_{Ref} \sim 1$. Therefore, with a perfectly anti-bunched light field as the reference field and an adjustable delay $T_o$, we may have a coincidence signal level that fully matches that of Michelson's stellar interferometry method \cite{mich}.

Since this technique relies on time-resolved coincidence measurement, that is, $T_R\ll T_c$, the bandwidth of the observed fields is limited by $1/T_R$. The current technology has $T_R\sim 10 ps$, which gives $T_c\sim 10 T_R = 100ps$ to satisfy $T_c\gg T_R$ and therefore a bandwidth of 10 GHz. This allows a maximum single-photon rate of $1/T_{Ref}\sim 1/T_R \sim 10^{10}/s$. The requirement of $T_{Ref}\sim T_R$ is to ensure that when an incoming photon is detected within the coincidence window of $T_R$, there is always a photon from reference fields for coincidence. Otherwise, the coincidence signal is dropped by a factor of $T_R/T_{Ref}$.

In conclusion, we extended the traditional phase insensitive Hanbury-Brown and Twiss interferometer to a phase sensitive one by mixing input fields at two locations with coherent fields and demonstrated the capability of measuring the complete complex coherence function of thermal fields. Although the scheme does not have the signal level as the stellar interferometer with entangled photons, it is based on coherent states and does not require a quantum network with entangled photons and is thus a trade-off between the availability of current technology and the future quantum technology. Moreover, the use of single-frequency coherent state as the reference light in our scheme leads to the requirement of time-resolved intensity correlation measurement. This may limit the bandwidth of the observed fields to that of the detectors. Nevertheless, it eliminates the temporal mode match issue between the input fields and the reference fields, as occurs in the pulsed case \cite{sm23}. Furthermore, our scheme also works for the input fields in coherent state, which is often the case when distant objects are actively illuminated by a CW laser or pulsed laser. In this case, the maximum visibility deduced from Eq.(\ref{As3}) will be 50\%. Therefore, our investigation is not only beneficial to astronomical applications, but also paves the way for developing new technology of remote sensing and interferometric imaging.


This work was supported in part by the National Natural
Science Foundation of China (Grant Nos. 12004279, and
12074283) and by City University of Hong Kong (Project No.9610522) and the General Research Fund from Hong Kong
Research Grants Council (No.11315822).

\end{document}